# A physically inspired model of Dip d792 and d1519 of the Kepler light curve seen at KIC8462852

*Eduard Heindl[1], Furtwangen University, Germany*

## Abstract

The star KIC 8462852 shows a very unusual and hard to comprehend light curve. The dip d792[2] absorbs 16% of the starlight. The light curve is unusually smooth but the very steep edges make it hard to find a simple natural explanation by covering due to comets or other well-known planetary objects. We describe a mathematical approximation to the light curve, which is motivated by a physically meaningful event of a large stellar beam which generates an orbiting cloud. The data might fit to the science fiction idea of star lifting, a mining technology that could extract star matter. We extend the model to d1519 and d1568 using multiple beams and get an encouraging result that fits essential parts of the dips but misses other parts of the measured flux. We recommend further exploration of this concept with refined models.

## 1. Introduction

The star KIC 8462852, known as Boyajian's Star, shows flux variations up to ~ 20% which lack a universally accepted interpretation. Since the first published article from Boyajian et al (Boyajian, 2016), there have been multiple interpretations of the strange dips. Boyajian favored a stream of large comets resulting in the measured signals.

A paper concerning the comets theory (Eva. H.L. Bodman, 2016) calculated the necessary comets to describe the dips around day 1506. Although the dips could be approximated by an event with multiple comets, Bodman explicitly mentions, that "the model does not explain the dip around day 800" (d792 event).

The dip d792 is about 16% deep and has an asymmetric shape. It lasts about 10 days when we include the very soft transition to normal flux. The central part of the dip is very steep and very hard to understand as a result from well-known orbital events like planet occultation or comets, as mentioned.

There have been multiple other objects and events included in the considerations, a comprehensive introduction is given by J. Wright (Wright, 2016). One interesting idea is an alien megastructure of unknown shape. The problem is that even a simple mega structure of orbiting objects hardly generates a dip with a shape as the Kepler space telescope observed at day 792 of the mission.

## 2. The Data

The data of the Kepler mission are publicly available at multiple sites[3] (Mikulski). The difficulty to use the datasets is that the Kepler team did some preprocessing to adjust for sensor drifting. Therefore, the data are not absolute values of the intensity. This does not influence the calculation within this article too much and is therefore ignored. An exact analysis of the absolute intensity of the light flux was published by Monet (Simon, 2016). The dip d792 under consideration falls in the first period of Monets analysis, were the global intensity of the star seems relative constant, with flux decay below 1% per year. The data analysis here covers only intervals over ten days, the resulting drift is therefore below the error level.

---

[1] Contact author hed@hs-furtwangen.de
[2] We name the dips with the letter d and a number for the day when the deepest dip of a main event appears.
[3] The dataset used here is from http://www.astro.caltech.edu/~btm/kic8462852/reduced_lc.txt, we used the 'normalized flux' column for calculation.

The flux of the star shows a 0.88 day period which is assumed to be the result of sun spots (Boyajian, 2016) or it could be the result of a star, that is by accident in the line of sight, as suggested by Makarov (Makarov, 1016). This may results in some different interpretation of the seen fluctuation at the edges of the dips.

## 3. The Model

The main aim of this paper is a physical model for the flux variation of the dip at day 792. The model is based on the idea, that a stream of matter leaves the star into space similar as observed in solar flares. The difference is, the stream of matter is quite high and lifts the matter into a stable orbit. It is not the aim of this paper to speculate for the mechanic of this event in detail. Astrophysical jets of matter are part of cosmic events and usually observed in combination with compact objects like neutron stars. These jets leave the object on the rotation axis and are therefore highly stable in the direction.

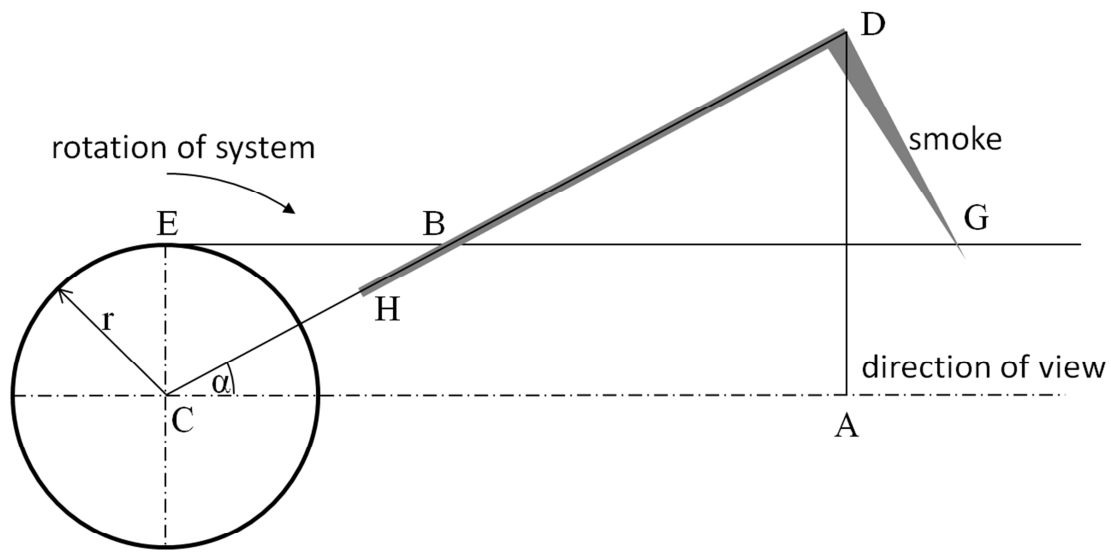

**Figure 1: Model of simulated object (not in scale)**

We assume a different type of jet; it starts at or near the equator of the star and is not stable in direction. The jet rotates with an orbital period in the range of years. Due to limited data, we have only one clean example of this dip type, and it might be that other deep dips result from a similar event. We don't know much about the rotation speed of the beam and the orbit. But due to the fact, that the Kepler telescope observed no other strong dip within 700 days, the period has to be at least 700 days. If the strong dips at the end of the Kepler mission are parts of a different event, then the period is at least 800 days.

The detail of the model used for the simulation is shown in Figure 1. The star with center C and radius *r* is the source of a visible beam HD. We assume, the beam leaves the star at the surface and the angle *α* is the angle to the direction of view AC by the Kepler space observatory.

Beginning with point H, the beam absorbs a part of the light from the star due to deionization and condensation. The beam comes to a stop at D. At this point, the beam does for some reason not continue and settles as "smoke" in an orbit in direction to point G.

The trail of smoke has a shape resulting in an effective optical density with exponential decay along the line DG, the triangle in the drawing is only a sketch of the real situation.

The beam and trail of smoke can be seen as an absorbing structure from the direction of view. To understand a dimming event by this model, we give a short description of the event. First the trail of smoke enters the line of sight EB near point G. Due to the exponential decay, there is no exact point in time of the beginning. The entrance of the smoke results in an exponential growth of the absorption. After a while, the beam at point H reaches the line of sight and starts to dim the flux further. The peak of the dimming is reached, when point D is in front of the star seen by the direction of view. When D leaves the stellar disk the absorption falls very rapid and the flux reaches normal level again.

There is no calculation concerning the intensity variation of the star disk itself, but we use the median of 2.5h by calculating the mean from five calculation intervals, each calculation interval covers 0.5h, to calculate a used intensity value that is displayed in the plots and is part of the error calculation.

## 3.1 Method of calculation

To calculate this dimming event scenario, we have to formulate the time dependent angle α in a useful way, we choose

$$\alpha(t) = ((BJD-2454833) - t_0)\omega \qquad (1)$$

where BJD (Barycentric Julian Date) is the time used by the Kepler data, $t_0$ describes the moment when D passes the line AC and ω is the rotation speed of the structure, this speed should not be confused with the rotation speed of the star, which might be much higher.

The visible part of the beam enters the field of view when

$$|\,|CH|\,\sin(\alpha)\,| < r \qquad (2)$$

The length of beam from H to B was computed using the geometric relation that the distance CB could be calculated by the known angle α at a time *t* and the radius *r* of the star.

$$|CB|\,\sin(\alpha) = r$$

We find

$$|CB| = r/\sin(\alpha)$$

Resulting in the visible length *l*

$$l = |\,r/\sin(\alpha) - |CH|\,| \qquad (3)$$

to include the proposed finite length of the whole beam HD, we have to check for entering the point D into the field of view, therefore an equation, similar to equation (2) was used

the point D enters the field of view when

$$|\,|CD|\,\sin(\alpha)\,| < r$$

In that case, the length of the visible beam is equal to the distance HD.

The flux of the star $f_0$ is reduced due to the absorption by the beam to $f_B$ in the line of sight, using the very simple model of an absorbing medium described by the absorption coefficient $\tau_B$.

$$f_B = f_0 \exp( l / \tau_B ) \qquad (4)$$

In addition to the flux reduction by the beam, we have to care about the smoke cloud DG in the orbit $|CD| = r_{orbit}$. The optical density $\tau_S$ of this cloud is by our definition dependent from the distance from the point D, the virtual source where it starts with the value $s_0$. Due to the fact, that we don't know absolute distances and for ease in computation, we describe the flux $f_S$ dependence from the angle $\alpha$

$$f_S = f_0 \, s_0 \exp( -\alpha / \tau_S ) \qquad (5)$$

in the case[4], that $\alpha < 0$ there should be no smoke and $f_S = f_0$.

The resulting flux $f(t)$ at time $t$ is then calculated by

$$f(t) = f_B f_S \qquad (6)$$

The model of a beam and cloud as described were computed for the same time intervals as the Kepler mission reports data. The calculation used the time named in the dataset and did in the first step not integrate over the measurement interval that was in reality about 20 minutes. This should be mentioned, because in the very steep edges of the dips even this can produce a significant error.

To reduce this and other effects of the unphysical point effects, for further processing the mean value $f_f(t)$ of five such calculated values were used, were Δt is the time interval between two measurements of the Kepler mission.

$$f_f(t) = \frac{1}{5} \sum_{k=-2}^{2} f(t + k\Delta t) \qquad (7)$$

### 3.2 Fitting the Data

The calculation was implemented in an Excel sheet that also contained the published data of the Kepler mission as described.

To fit the calculated and the measured data, the mean square error over the time interval of study was calculated. The optimization started manually by variation of the free parameters presented in table 1.

The final optimization for every parameter used the built in optimizer in Excel. The fact, that all parameters were allowed to change, resulted in a radius of the star that is not equal to one, but this numeric value should not imply any size, due to the unit free calculation in the whole simulation, except the time, that is coupled to mission time as described.

It is not a trivial problem, what exactly should be optimized. During the procedure of optimization it was obvious, that a perfect shape at the peak of the dip was not the very best visible fit at the edges, although the difference was not very high.

The result presented for the different dips were in this way different in the optimization, in the case of d792, the whole interval was used and the mean square error was minimized. In the case of d1519 and d1568 the fit was completely manually optimized for best visual and plausible results.

---

[4] This holds only for |angles| < π, all calculations used only small angles in this sense.

**Table 1 Used parameters**

| parameter | D792 | D1519 | D1568 |
|---|---|---|---|
| $t_0$ [d] | 792.7260619 | 1518.81 | 1567 |
| $\omega$ [1/d] | 1.00E-02 | 1.00E-02 | 1.00E-02 |
| $\tau_B$ | 6.30E+03 | 6.30E+03 | 6.30E+03 |
| $s_0$ [rad] | 0.021290177 | 0.021290177 | 0.012129018 |
| $\tau_S$ | 0.017269375 | 0.016666667 | 0.012360027 |
| $r_0$ | 0.925043748 | 0.925043748 | 0.925043748 |
| $h_0$ | 32.48115357 | 40.8065473 | 138.7565622 |
| $r_{orbit}$ [$r_0$] | 955.015 | 955.015 | 1259.984 |
| **parameters for multi-dip events** | | | |
| $t_1$ [d] | | 1519.499 | 1567.59 |
| $t_2$ [d] | | 1519.74 | 1568.49 |
| $b_0$ | | 0.29 | 0.047 |
| $b_1$ | | 0.93 | 0.123 |
| $b_2$ | | 0.87 | 0.413 |

## 4. Result for d792

The result is displayed in Figure 2, a first view shows an astonishing match between calculation and measurement. The only area, were the measured flux and the calculated flux are near 1% of the flux is near the deep dip. But this fluctuation may result from the very simple model of the beam to smoke conversion process. It was only assumed, that the smoke starts at t=792.515568, without keeping care how the details are. It seems possible to reduce this by introducing a model for that conversion, but this is not the aim of this paper.

Another visible error is in the time interval 793 to 794, where the calculated value was a little bit overestimating the measured value. During the optimization of the parameters it was visible that minor changes of the distance CH, the place where the condensation happens, can reduce this error. A model using a continuous model of this process could be useful at his part of the approximation of the flux.

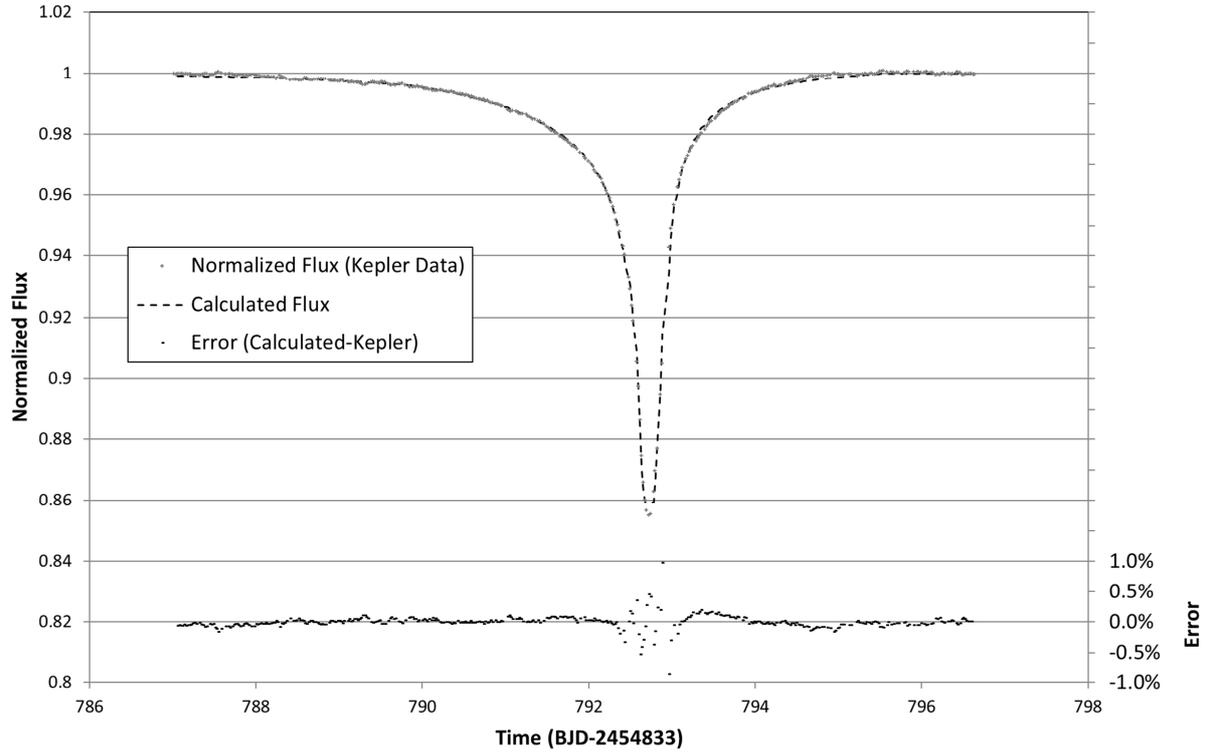

**Figure 2: Dip d792 measured data, simulated data and error**

Very interesting is the very small error at the left part of the plot, supporting the exponential model of the smoke density. It might be part of a continuous growth of the particles, and should be further compared to well-known physical models of nebular dynamics.

## 5.  Extending the model to dip d1519 and d1568

The success of the model for dip d792 suggests it should also be applied to other dips, in particular to the dip formation around d1519 and d1568. Within this formation, at least three dips similar to the d792 dip can be identified. The analysis uses the basic concept as described in part 3 only minor changes in the parameters of the beam and smoke system were introduced as noted in Table 1.

In detail, the time of the first dip in each case called $t_0$ was set to the actual first dip in the event. The following variables were kept with the same value:

Rotation speed $\omega$, absorption parameter of the beam $\tau_B$, starting density of the smoke cloud $s_0$, orbital radius $r_{orbit}$ of point D and of course the star radius $r_0$.

To model the three main dips the model of the obscuring system is repeated three times and every system is shifted to the position by a time shift $t_i$ to fit the concerned dip. Then the absorption of the obscuring system is modified by a linear factor $b_i$, to match the measured dips and lead to a low error.

The resulting transmission $T_r(t)$ is then calculated using the equation

$$T_r(t) = \prod_{i=0}^{2} b_i\, T_b(t - t_i)\, T_s(t - t_i) \qquad (8)$$

To fit the parameters to the measured values of the flux, only parts of the observation time were used.

## 5.1 Dips d1519

The result for the series of d1519 is shown in Figure 3

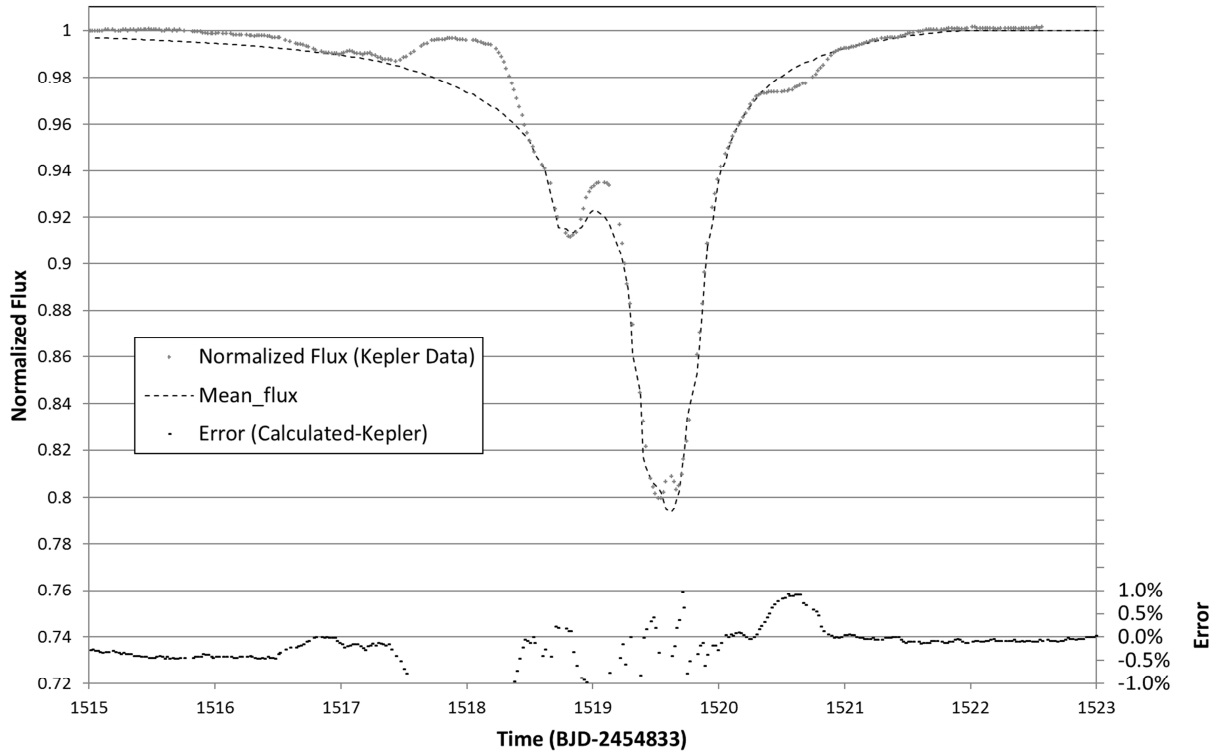

**Figure 3: Dips D1519**

The result is on the first view not very convincing. But we should have a closer look to some interesting details of the calculated flux and the measured flux.

The central very deep dip, more than 20%, could be reproduced by two beams of the type d792, all parameters see Table 1. The error is within 1% of the absolute flux in this area t = 1519 till the end at t = 1523 and may be a signal to go deeper into this model.

Especially the right edge has only one bumper, which is apparent, the result of another object in the line of sight. Excluding this object, the error is below 0.1% and is within the noise value of the measured light.

At the left side, the used model overestimates the absorption. There can be multiple reasons; two of them should be noted.

One reason could be that the ejection of material from the star was not continuous and as a result there are gaps in the smoke cloud

Another more exotic reason might be, that the smoke is raw material for a civilization and is partly already harvested, resulting in the reduced flux seen from day 1515 to 1516.5 and day 1517.5 till day 1518.5.

## 5.2 Dips d1568

The last set of deep dips was observed by the Kepler mission around day 1568. This set of dips looks in some parts similar to the d1519 set but has not the same deep flux reduction as seen in dip d1519.

To understand the model of a beam and the absorption better, we also applied the concept and reached the result displayed in Figure 4.

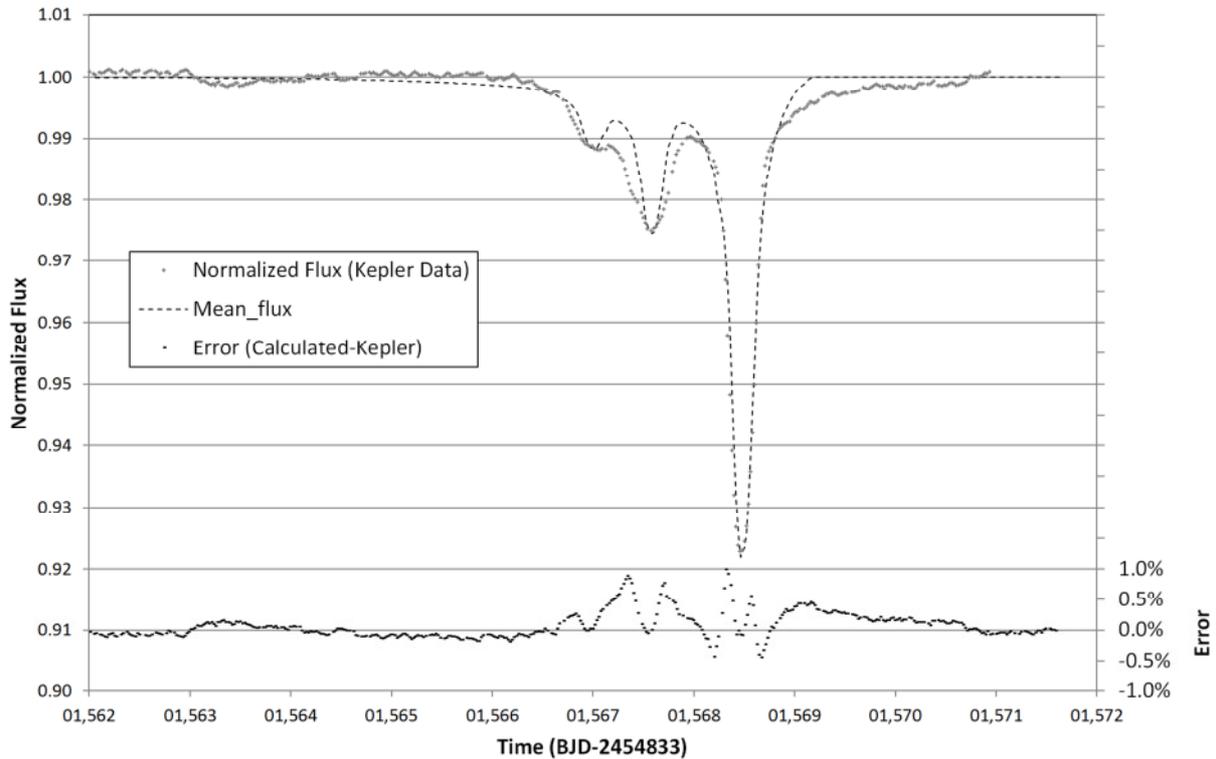

**Figure 4: Dips around day 1568**

To reach an acceptable result, the parameters of the beams have to be changed sharply for this case. Especially the height $h_0$ were the beam gets optical dense was lifted to 138.75 star radii. This leads to a result, that seems to give the shape of d1568 a reasonable low error, but it is hard to find therefore a simple physical reason.

One reason might be that the material ejection from the sun stopped suddenly and we see only the remaining material entering the outer part of the beam trajectory.

The smaller dips were included with the same parameters and were shifted to their position by $t_0$ and $t_1$ but did not cover the observed flux reduction.

It would be possible; to change all available parameters for this case to reduce the error, but there should be good physical reasons due to the circumstances, that we did not try to find a new type of spline but an interesting explanatory model for the observed flux variations.

5.3   Clean Mirror Effect

There is another effect near to the big dips around d1519 that should be noted. If we calculate the moving average over a period of 83 intervals, we see only significant changes in the flux as shown in Figure 5.

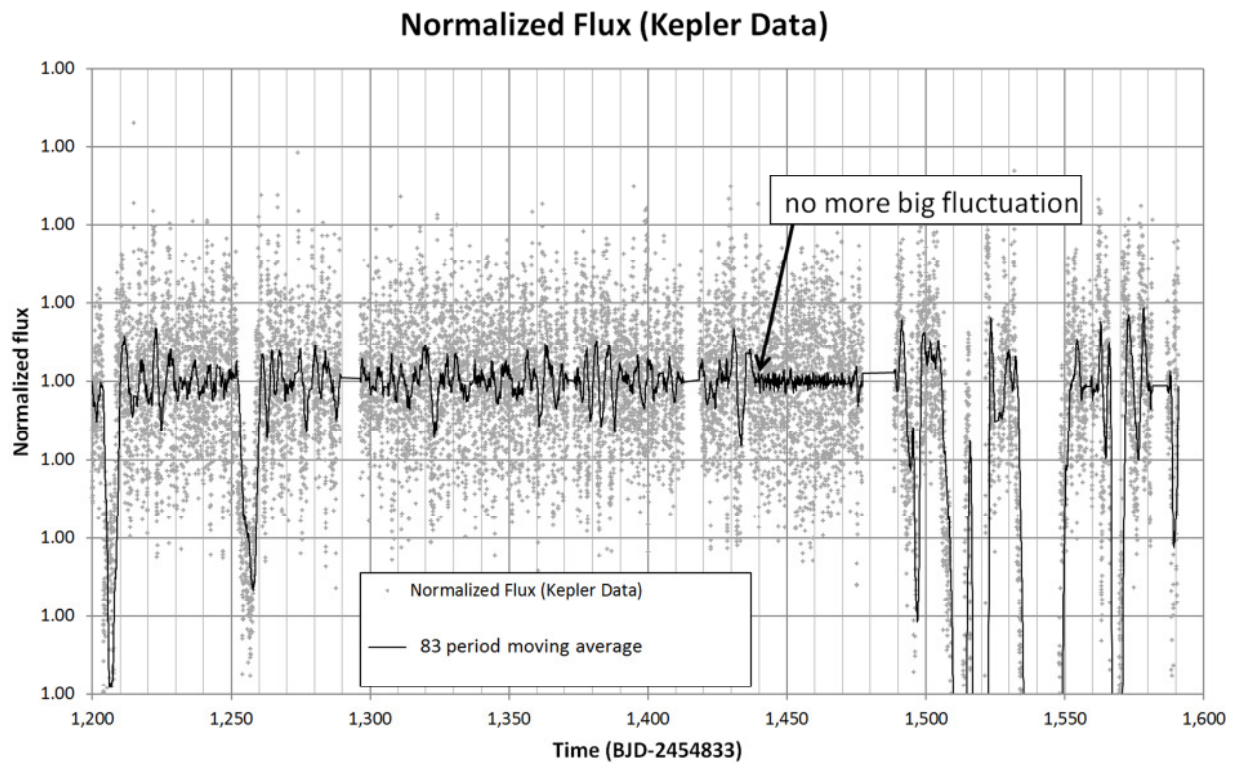

**Figure 5: Moving average**

The interesting signature is a just complete drop of the 'noise' at day 1440, 60 days before the very large series of dips starts. By visual inspection, it seems a highly significant change in the measured signal. As long as this is not an artefact from Kepler, and we don't think it is one, then it might be correlated to the following sequence of dips and needs an explanation.

If there is some type of mining business going on and a smoke cloud is generated, as proposed in this paper, it would be very stupid to exhibit different equipment in the orbit, like mirrors to the dust. So there may be a limited area where no large equipment should go. And then we see this very low noise signal as presented.

## 6. Discussion

The Boyaijans Star is still a big puzzle to investigate, but we hope, this paper gives a fresh impetus to analyse the data open minded for the unexpected ETI source of flux variations.

We have seen, that at least d792 is very precisely modeled by a very simple model with a low number of free parameters. The remaining error might be the result of oversimplification, but it was not the aim of the paper to model a perfect spline with dozens of free parameters, instead to find a physical model for the dips.

The result is a beam, starting at the surface of the star and radiating matter up to an orbit 1000 times higher than the stellar radius, comparable to the Jupiter orbit, this seems to be a physical possible system. Including a smoke cloud in the orbit is a logic result of the beam.

The physical values of energy and matter flows have not been calculated due to many open questions, but could be nailed down, when further data, including spectral data of the absorption of the beam, could be generated.

The system may be the result of a process called star lifting. A concept, were an ETI uses the matter of its star for mining activities. Therefore the other signals of the Kepler mission should be analyzed to find further hints, as in a first example was shown with the missing signal of large objects before the very strong dips appeared.

We thank the community at https://www.reddit.com/r/KIC8462852/ for fruitful discussions.